\documentclass[aps,preprint,preprintnumbers,eqsecnum,prb]{revtex4}
\usepackage{amsfonts}
\usepackage{makeidx}
\usepackage{graphicx}
\usepackage{amsmath,amsbsy,amssymb,graphics}

\begin{document}

\title{Peculiar Band Gap Structure of Graphene Nanoribbons}
\author{Motohiko Ezawa}
\affiliation{Department of Physics, University of Tokyo, Hongo 7-3-1, 113-0033, Japan }

\pacs{73.22.-f,78.67.Pt,78.66.Sq\hskip1cm \textbf{DOI:} DOI: 10.1002/pssc.200790003; DOI: 10.1002/pssc.200673205}

\begin{abstract}
Graphene nanoribbons are quasi-one-dimensional meterials with finite width.
Characterizing a wide class of nanoribbons by edge shape and width, we make
a systematic analysis of their electronic properties. The band gap structure
of nanoribbons is shown to exhibit a valley structure with stream-like
sequences of metallic or almost metallic nanoribbons. Among them, all zigzag
nanoribbons are metallic, and armchair nanoribbons are metallic by period of
3. We find that these stream-like sequences correspond to equi-width curves,
and that the band gap of chiral and armchair nanoribbons oscillate as a
function of the width. Furthermore a possible application of nanoribbons to
nanoelectronics is discussed. \par\textbf{DOI: 10.1002/pssc.200790003; 
DOI: 10.1002/pssc.200673205}
\end{abstract}

\preprint{Physica Status Solidi (c) 4, No.2, 489 (2007); Cover Picture of this Issue}

\maketitle

\section{Introduction}

Electronic properties of carbon systems are dramatically depend on their
geometry and size. In particular, intensive research has been made on carbon
nanotubes in the last decade. The large interest centers their peculiar
electronic properties inherent to quasi-one-dimensional systems.

A similarly fascinating carbon system is a ribbon-like stripe of a graphite
sheet, which is named graphene nanoribbons or carbon nanoribbons\cite%
{Ezawa,Nakada}. Recent experimental developments\cite{Novoselov} enable us
to isolate a graphene, which is a monolayer graphite. Nanoribbons have a
higher variety than nanotubes because of the existence of edges. Though
quite attractive materials, their too rich variety has made it difficult to
carry out a systematic analysis of nanoribbons.

\begin{figure}[!tbh]
\includegraphics[width=0.6\textwidth]{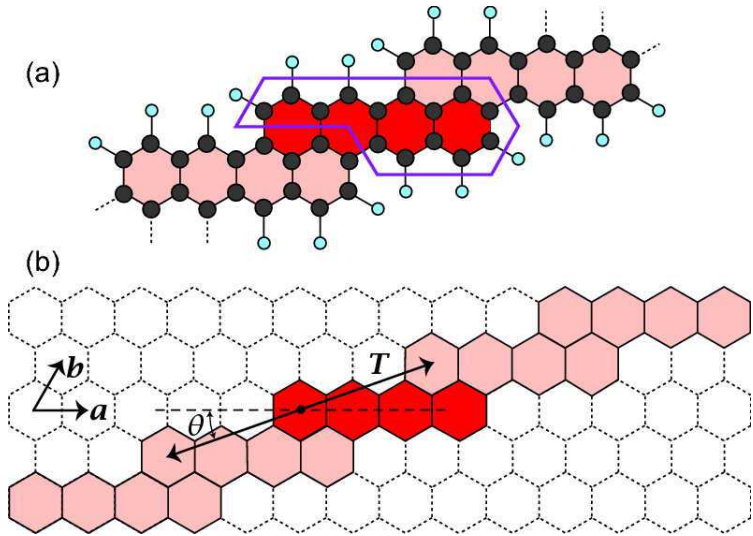}
\caption{(a) A typical structure of nanoribbons. A black circle stands for
a carbon atom with one $\protect\pi $ electron, while a blue circle for a
different atom such as a hydrogen.\ A closed area represents a unit cell. It
is possible to regard the lattice made of black circles as a part of a
honeycomb lattice.\ (b) A nanoribon is constructed from a chain of $m$
connected carbon hexagons, as depicted in red, and by translating this
chain by the translational vector $\mathbf{T}=\pm q\mathbf{a}+\mathbf{b}$
many times, as depicted in pink, where $q<m$. A nanoribbon is indexed
by a set of two integers $\langle p,q\rangle $\ with $p=m-q$. Here we have
taken $m=4$, $q=2$, $p=2$. }
\end{figure}

In this work, characterizing a wide class of nanoribbons by a set of two
integers $\langle p,q\rangle $, we present a systematic analysis of their
electronic property in parallel to that of nanotubes. They are shown to
exhibit a rich variety of band gaps, from metals to typical semiconductors.
We reveal that there exist sequences of metallic or almost metallic
nanoribbons which look like streams in valley made of semiconductors. They
approach equi-width curves for wide nanoribbons. We also find a peculiar
dependence of the electronic property of nanoribbons on the width $w$.
Furthermore, we point out that the variety of nanoribbons make them
promising candidates for electronic device applications.

\section{Classification and Electronic Structures of Graphene Nanoribbons}

The classification rule for the nanotubes says that it is metallic when $%
n_{1}-n_{2}$ is an integer multiple of $3$, and otherwise semiconducting,
where $\left( n_{1},n_{2}\right) $ is a chiral vector of the nanotube.
Suggested by this classification, as we illustrate in Fig.1, we consider a
wide class of nanoribbons indexed by a set of two integers $\langle
p,q\rangle $ representing edge shape and width. Nanoribbons with $q=0$
possess zigzag edges, nanoribbons with $q=1$ possess armchair edges and the
other nanoribbons possess chiral edges.

\begin{figure}[!tbh]
\includegraphics[width=0.7\textwidth]{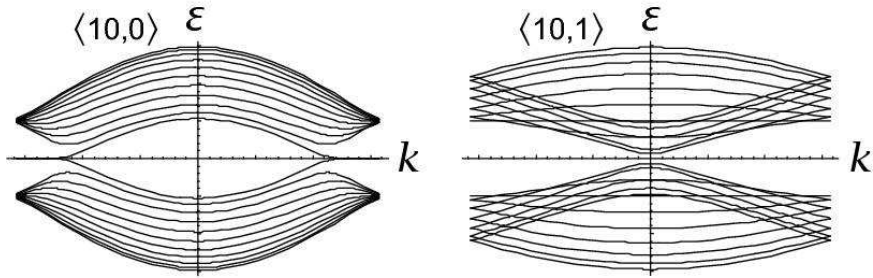}
\caption{{}The band structure of nanoribbons. The horizontal axis is the crystal momentum $k$, $-\pi < k <\pi$, 
while the vertical axis is the energy $\epsilon$, $-3|t| <\epsilon < 3|t|$ with $|t|=3.033$eV. The band structure depends strongly on the 
index $q$ for fixed $p$, but depends on the index $p$ only weakly for fixed $q$.}
\end{figure}

\begin{figure}[!tbh]
\includegraphics[width=0.45\textwidth]{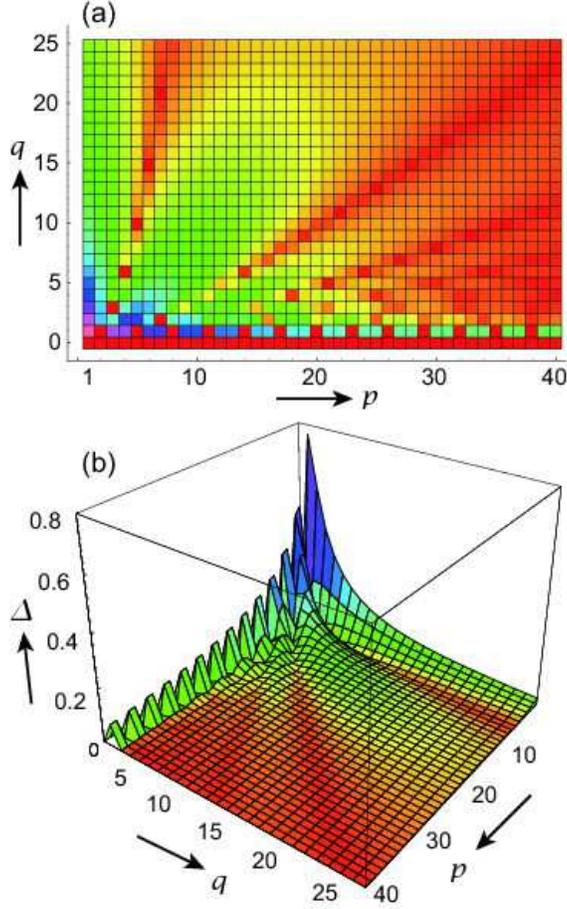}
\caption{{}The band gap structure of nanoribbons. (a) The horizontal and
vertical axes represent the indices $p$ and $q$, respectively. Magnitudes of
band gaps are represented by colored squares. Redish (blue) gray squares
represent small (large) gap semiconductors. Especially, metallic states are
represented by red squares. (b) A bird's eye view. The vertical axis is
the energy gap $\Delta $ in unit of $|t|=3.033$eV. Nanoribbons make a valley
structure with stream-like sequences of metallic points in the $pq$ plane. We 
observe clearly three emergence patterns of metallic points: (a) All zigzag 
nanoribbons are metallic points. (b) Armchair nanoribbons with $p=2,5,8,11$,
is metallic (c) Several sequences of metallic points on "streams" in valleys. 
In particular, nanoribbons are metallic on the points $\langle p,q\rangle $ with integers $p$ and $q$ with $
q=p(p-1)/2$. They constitute the principal sequence of metallic points. 
Several sequences of metallic or almost metallic points are found in 
the valley of semiconducting nanoribbons. These sequences correspond 
to equi-width curves indexed by w as $p=-q+(w/2)\sqrt{(3(2q+1)^2+9)}$.}
\end{figure}

We carry out a systematic analysis of the electronic property of
nanoribbons. It is well known that the electronic properties of graphene and
nanotube are well explained by taking the nearest neighbor tight banding
model. The analysis of nanoribbons can be done in a similar way. The
tight-binding Hamiltonian is defined by%
\begin{equation}
H=\sum_{i}\varepsilon _{i}c_{i}^{\dagger }c_{i}+\sum_{\left\langle
i,j\right\rangle }t_{ij}c_{i}^{\dagger }c_{j},  \label{HamilTB}
\end{equation}%
where $\varepsilon _{i}$ is the site energy, $t_{ij}$ is the transfer
energy, and $c_{i}^{\dagger }$ is the creation operator of the $\pi $
electron at the site $i$. The summation is taken over the nearest neighbor
sites $\left\langle i,j\right\rangle $. Carbon nanotubes are regarded as a
periodic-boundary-condition problem, while carbon (graphene) nanoribbons are
as a fixed-boundary-condition problem imposed on graphene.

\begin{figure}[!tbh]
\includegraphics[width=0.6\textwidth]{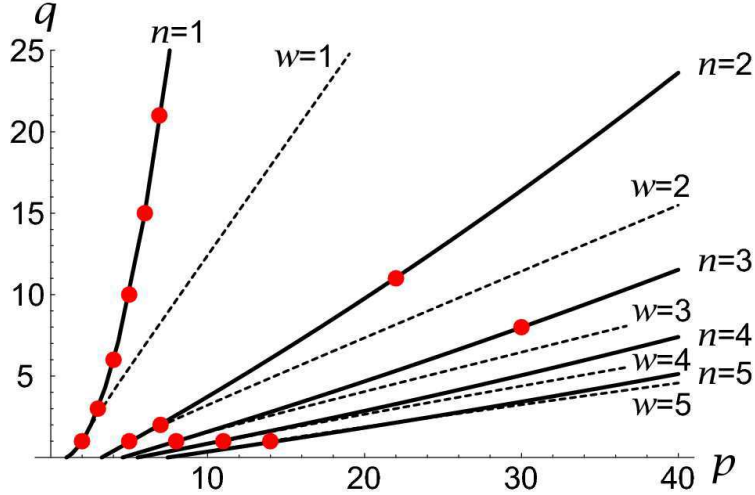}
\caption{{}Illustration of metallic points, sequences and equi-width curves.
Metallic points are denoted by red circles. Solid curves represent
sequences of metallic or almost metallic points, while dotted curves represent the points $\langle
p,q\rangle $ possessing the same width. The width w is defined as
$w=2(p+q)/\sqrt{(3(2q+1)^2+9)}$.
The $n$-th sequnce is tangent to the
equi-width curve with $w=n$ at $q=1$. These two curves become almost
identical for sufficiently wide nanoribbons.}
\end{figure}

\begin{figure}[!tbh]
\includegraphics[width=0.6\textwidth]{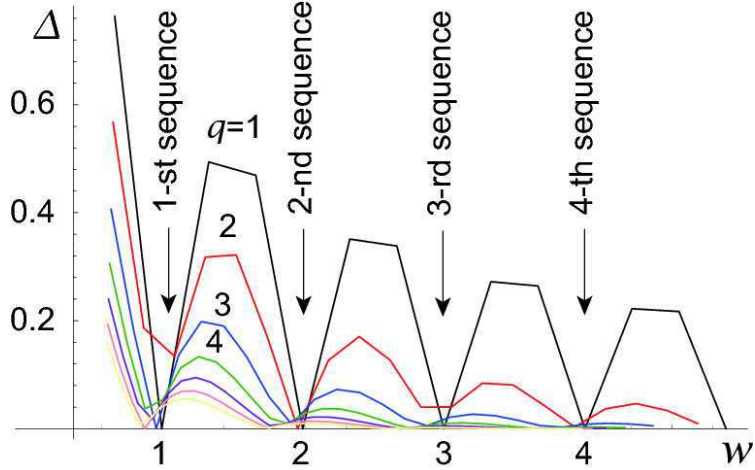}
\caption{{}The band gaps $\Delta$ in unit of $|t|$=3.033eV as a function of the width $w$. 
They oscillate and take local minima almost at the same values of the width $w$ for any $q.$
Their envelope decrease inversely against $w$.}
\end{figure}

The band gap of nanoribbons is calculated for each point $\langle p,q\rangle 
$, as in Fig.2. Collecting all these results, we display the band gap
structure in Fig.3. Nanoribbons exhibit a variety of properties in
electronic conduction, from metals to typical semiconductors. It is
remarkable that it makes a valley structure with stream-like sequences of
metallic or almost metallic nanoribbons (Fig.3). It is observed that
nanoribbons indexed by $\langle p,0\rangle $ are metallic for all $p$, which
are in the polyacene series with zigzag edges. Nanoribbons indexed by $%
\langle p,1\rangle $ with $p=2,5,8,11,\cdots $ are found to be also
metallic, which have armchair edges. This series has period $3$, as is a
reminiscence of the classification rules familiar for nanotubes.

We point out a peculiar dependence of the electronic property of nanoribbons
on the width $w$, where%
\begin{equation}
w=\frac{2\left( p+q\right) }{\sqrt{3\left( 2q+1\right) ^{2}+9}}.
\label{WidthParam}
\end{equation}%
We extract the stream curves out of Fig.3, and draw them in Fig.4. On this
figure we also present equi-width curves. It is remarkable that these
sequences and equi-width curves become almost identical for wide
nanoribbons, as is clear in Fig.4. We have depicted the band gap as a
function of the width $w$ for each fixed $q$ in Fig.5. The band gaps
oscillate and the envelope of the band gap decreases inversely against $w$.

\section{Van Hove singularities}

\begin{figure}[!tbh]
\includegraphics[width=0.7\textwidth]{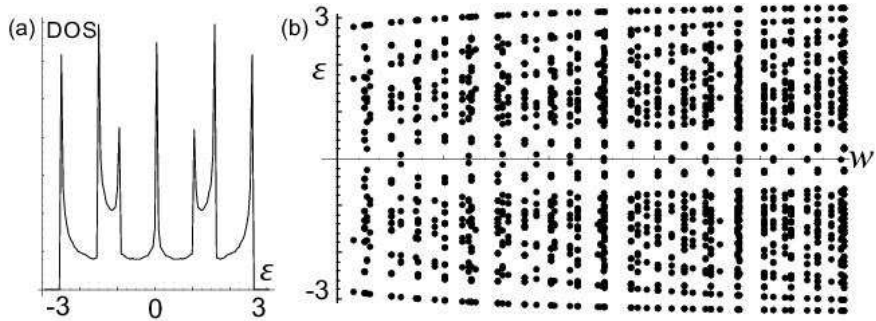}
\caption{(a) The density of state (DOS) of the $\langle 1,0\rangle$ nanoribbon. 
(b) Plot of van Hove singularities in the $w$-$\varepsilon$ plane. 
For a given $\langle p,q\rangle$ nanoribbon, we calculate the width w and the energies 
$\varepsilon$ at which van-Hove singularities develop. We have plotted the points ($w$,$\varepsilon$) 
for $q=0,1,2,3,4$ and for all $p$ in the region $w<3$. A stripe pattern is manifest.}
\label{EdgeBandE}
\end{figure}

We calculate the energies $\varepsilon $ at which van-Hove singularities
develop due to the local band flatness at $k=0$ for various $\left\langle
p,q\right\rangle $ nanoribbons [Fig.6(a)]. Note that the optical absorption
is dominant at $k=0$ because the dispersion relation $\varepsilon =ck$ with $%
c$ the light velocity. On the other hand the width $w$ is determined by $p$
and $q$ as in (\ref{WidthParam}). We show the energy $\varepsilon $ of this
peak as a function of $w$ in Fig.6. A peculiar stripe pattern\ is manifest
there. In particular, the maximum and minimum values take almost the same
values $\pm 3\left\vert t\right\vert $, reflecting the electronic property
of a graphite. The fact that there are on smooth curves present another
justification to call $w$ the width of nanoribbon.

\section{Devices}

\begin{figure}[!tbh]
\includegraphics[width=0.9\textwidth]{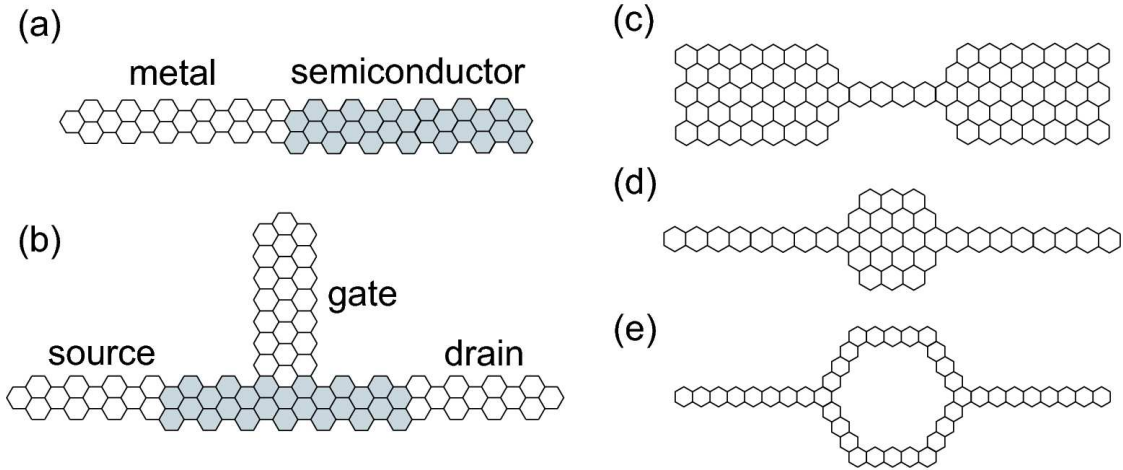}
\caption{{}Several possible nanoribbon devices.
(a) Metal-semiconductor hetro-junction of nanoribbons with different indices (Schottky diode).
(b) Hetro-junction of nanoribbons with transistor configuration (Schottky gate field
effect transistor). (c) Point contact. (d) Quantum
dot. (e) Aharonov-Bohm ring.}
\end{figure}

We have revealed a rich variety of band gaps in graphene nanoribbons. They
are either quasi-one-dimensional metals or semiconductors depending on their
edge shape and width. Graphene nanoribbons could be promising candidates of
molecule devices, similarly to nanotubes\cite{Yao}, because of ballistic
transport at room temperature.

We mention the merits of nanoribbons in comparison to nanotubes. Two
nanoribbon segments with different atomic and electronic structures can be
seamlessly fused together to create intramolecular metal-metal,
metal-semiconductor, or semiconductor-semiconductor junctions without
introducing a pentagon and a heptagon into the hexagonal carbon lattice.
Diodes or transistors could be made of nanoribbons, as illustrated in Fig.7.
For instance, a metal-semiconductor junction makes Schottky barrier and may
behave like a Schottky diode. Similarly Schottky gate field-effect
transistor, point contact, quantum dot and Aharonov-Bohm ring may be
realized. We might even design a complex of electronic circuits by etching a
monolayer graphite in future. It is interesting problems to calculate
electronic characteristics of various junctions, which will be studied
elsewhere.


\begin{thebibliography}{1}
\bibitem[1]{Ezawa} M.~Ezawa, Phys.~Rev.~B \textbf{73}, 045432 (2006).

\bibitem[2]{Nakada} K.~Nakada, M.~Fujita, G.~Dresselhaus and M.~S.
Dresselhaus, Phys.~Rev.~B \textbf{54} 17954 (1996).

\bibitem[3]{Novoselov} K.~S.~Novoselov \textit{et al}. Nature, 438 197
(2005).

\bibitem[4]{Yao} Z.~Yao, H.~W.~Ch.~Postma, L.~Balents and C.~Dekker, Nature, 
\textbf{402}, 273 (1999).
\end{thebibliography}
\end{document}